\documentstyle[12pt]{article}
\newcommand{\beq}{\begin{equation}}
\newcommand{\eeq}{\end{equation}}
\newcommand{\bea}{\begin{eqnarray}}
\newcommand{\eea}{\end{eqnarray}}

\begin{document}
\setcounter{page}{0}
\topmargin 0pt
\oddsidemargin 5mm
\renewcommand{\thefootnote}{\fnsymbol{footnote}}
\newpage
\setcounter{page}{0}
\begin{titlepage}

\begin{flushright}
OU-TP-98-41P \\
QMW-PH-98-35 \\
{\bf hep-th/9807223}\\
 {\it July 1998}
\end{flushright}
\vspace{0.5cm}
\begin{center}
{\Large {\sc  GRAVITATIONALLY DRESSED RG FLOWS AND
  ZIGZAG-INVARIANT STRINGS}} 
\\
\vspace{1.8cm}
\vspace{0.5cm}
{\sc Ian. I. Kogan$~^{1,}$\footnote{e-mail:
i.kogan1@physics.ox.ac.uk} and O. A.
Soloviev$~^{2,}$ 
\footnote{e-mail: O.A.Soloviev@QMW.AC.UK}}  \\
\vspace{0.5cm}
{\em$~^1$Theoretical Physics, Department of Physics, University of Oxford\\
1 Keble Road, Oxford, OX1 3NP, United Kingdom}\\
{\em$~^2$ Physics Department, Queen Mary and Westfield College, \\
Mile End Road, London E1 4NS, United Kingdom}\\
\vspace{0.5cm}
\renewcommand{\thefootnote}{\arabic{footnote}}
\setcounter{footnote}{0}
\begin{abstract}
{We propose a world-sheet realization of the zigzag-invariant bosonic
and fermionic strings as  a perturbed Wess-Zumino-Novikov-Witten model 
at large negative level $k$ on a group manifold $G$ coupled to 2D  
gravity. In the  large $k$ limit  the zigzag symmetry can be obtained 
as a result of a self-consistent solution of the gravitationally  dressed 
RG equation. The only solution  found for simple group is $G=SL(2)$. More 
general target-space geometries  can be obtained  via tensoring of various 
cosets based on $SL(2)$. In the supersymmetric case the zigzag symmetry 
fixes  the maximal target-space  dimension of the confining  fermionic string 
to be seven.}
\end{abstract}
\vspace{0.5cm}
 \end{center}
\end{titlepage}
\newpage

\section{Introduction}
\setcounter{equation}{0}

A hunt for the string theory formulation of QCD has,
 recently, entered a very intriguing  stage. 
A year ago Polyakov put forward a new idea about how to get a string 
 description of a gauge field theory without running into a problem of
 having spin-2 massless particles in the 
open string spectrum \cite{Polyakov}. 
 The main ingredient of this string formulation is a new 
symmetry on the world-sheet which he called zigzag symmetry. 
This is nothing but a symmetry under  orientation changing 
 world-sheet metric diffeomorphisms, $det(\partial x' / \partial x) <
 0$.  The power of the
zigzag  symmetry is that it singles out only vector states of the open
string in  the (D-dimensional) domain of the target-space to which 
 Wilson loops of the confining string are attached. In other words, 
 the zigzag symmetry gives rise to string models without gravity 
in the space-time.

A world-sheet description of the confining string appears to be fairly
intricate \cite{Polyakov},\cite{Alvarez}. 
It looks to be a certain non-critical string whose  
 world-sheet action is given by \cite{Polyakov}
\begin{eqnarray}
S=\int d^2\xi\;\left[(\partial\phi)^2~+~a^2(\phi)
 (\partial x)^2~+~\Phi(\phi)R^{(2)}\sqrt g\right]
\label{ansatz}
\end{eqnarray}
 where we omit possible antisymmetric fields, like $B_{\mu\nu}$ or
 Ramond-Ramond fields. 
 Here  $\phi$ is the Liouville field of 2D gravity, 
$x^\mu$ ($\mu$ runs from 0 to $D-1$) are 
 coordinates of the confining string, $\Phi$ is the dilaton field,
 $R^{(2)}$ is the curvature 
of the world-sheet and $a(\phi)$ is the running string tension.
 The zigzag symmetry requires
the existence of a certain value of the Liouville field $\phi^{*}$ such that
\begin{equation}
a(\phi^{*})=0
\label{boundary}
\end{equation}
At the given point, the string  representation of the Wilson 
loop is determined by the world-sheet action for antisymmetric
 $B_{\mu\nu}$ and/or R-R fields \cite{Polyakov}. One can think about
 the surface of a zero string tension as a horizon in a curved
 $D+1$-dimensional space.

According to Polyakov's ansatz, the colour-electric flux of the
D-dimensional gauge field  theory propagates in at least D+1
dimensions, so that the
D-dimensional gauge theory emerges on the boundary of a
(D+1)-dimensional  space-time (with negative curvature). 
 A very similar geometry of a target-space 
 was recently used in a recent AdS/CFT construction for $N=4$ SUSY
 gauge theories \cite{Maldacena},\cite{Gubser},\cite{Witten}.  

The aim of the present paper is to take a step further towards a
 concrete world-sheet formulation of a string model 
which is invariant under the zigzag symmetry. To find relevant 
 geometry  one has to study the beta-function for $a(\phi)$.
 The major unsolved  problem is
 to go beyond one-loop approximation  for beta function which is
invalid for  asymptotically small  $a(\phi)$  and large negative
  curvature  near the
horizon. In this paper we shall approach this problem by
 studying the gravitationally dressed RG flow \cite{Schmidhuber},
\cite{Klebanov} describing the deformation of a non-compact WZNW
 model by the kinetic term operator in the large $k$ limit.
 We shall look for  group manifolds such that  a  dressed RG
  flow  has a  nontrivial IR  fixed point at which the kinetic term
  is canceled, i.e. at this point the target space metric is zero and
 we have pure Wess-Zumino term interacting with gravity. 
 It is amusing that solution of this problem exists !
 
\section{Formulation of the bosonic zigzag-invariant string}
\setcounter{equation}{0}

The main objective of the world-sheet formulation of the
zigzag-invariant  string is to obtain 
a conformal field theory which has a sector with the running
 coupling constant in front of 
the kinetic terms. Our first step will be to consider a certain
 non-conformal two-dimensional 
model with the running constant in front of the kinetic term.
 Such a theory has been 
discussed in \cite{Soloviev1}. It is a (non-unitary)
 Wess-Zumino-Novikov-Witten (WZNW) model 
perturbed by its kinetic term. In two-dimensional complex coordinates,
the corresponding action is written as follows
\begin{equation}
S_\epsilon=S_{WZNW}(G,k)~-~\epsilon\int d^2z~O(z,\bar z).
\label{perturbation}
\end{equation}
Here $\epsilon$ is a  coupling constant,
\begin{eqnarray}
S_{WZNW}(G,k) ={k\over 4\pi}\int d^2 z 
Tr\left(g^{-1}\partial_zg\cdot g^{-1}\partial_{\bar z}g\right) +
 \Gamma(G,k)
\label{WZNW} 
\end{eqnarray}
 where the last term is a Wess-Zumino term
\begin{eqnarray}
\Gamma(G,k)={k\over 12\pi}\int_M d^3 x\,\,\epsilon^{ijk}
Tr g^{-1}\partial_ig\cdot
    g^{-1}\partial_jg\cdot g^{-1}\partial_kg,
\end{eqnarray}
and
\begin{equation}
O(z,\bar z)={1\over c_V(G)}J^a\bar J^b\phi^{ab},
\label{O}
\end{equation}
where we use the following notations
\begin{eqnarray}
J\equiv J^at^a=-{k\over2}\partial gg^{-1},~~~~~
\bar J \equiv \bar J^at^a=-{k\over2}g^{-1}\bar\partial g,
\label{currents}
\end{eqnarray}
and also
\begin{eqnarray}
\phi^{ab} = \mbox{Tr}(g^{-1}t^agt^b),~~~~~
c_V(G) = -{f^{ac}_df^{bd}_c\eta_{ab}\over\dim G}.
\label{notations}
\end{eqnarray}

The operator $O$ in eq.(\ref{notations}) has the following conformal 
dimension 
\cite{Knizhnik}
\begin{equation}
\Delta_0=1~+~{c_V(G)\over k + c_V(G)},\label{Delta}\end{equation}
which is positive and less than one if
\begin{equation}
k<-2c_V(G).\label{k}
\end{equation}
This condition is important, since we want $O$ to be a relevant
operator, so that the 
conformal point $\epsilon_0=0$ is unstable.
 The negative $k$ means that we are dealing with a non-unitary
 WZNW model: there will be states with negative norms in the
spectrum.  It is important that the operator $O$ 
itself generates a unitary highest weight representation of the 
Virasoro algebra, because we  choose $\Delta_0$ be positive.
 Later we shall show that only the unitary subsector of the 
given theory will play a role.

Another important property of the operator $O$ is that it
 obeys the following fusion rule 
\cite{Soloviev2}
\begin{equation}
O\cdot O=[{\bf1}]~+~[O]~+~...,\label{fusion}\end{equation}
where the square brackets denote the contributions of $O$ and the 
identity operator $\bf1$ 
and their descendants, whereas dots stand for operators 
with conformal dimensions greater 
than one. The given OPE implies that the perturbed CFT is  renormalizable.

Since the critical point  $\epsilon_0=0$ is unstable, 
the theory (\ref{perturbation}) will flow towards some IR fixed point
 or to a massive phase. 
The corresponding renormalization group beta function of 
the coupling constant $\epsilon$ is 
given by \cite{zam}, \cite{Cardy}
\begin{equation}
\beta\equiv{\mbox{d}\epsilon\over\mbox{d} \ln \Lambda}=
(2-2\Delta_0)\epsilon~-~\pi C\epsilon^2~+~...
\label{beta}\end{equation}
where $\Lambda$ is the ultraviolet cutoff.
Here the (three-point function) coefficient $C$ is normalized to 1,
 due to the factor 
$1/c_V(G)$ in the definition of $O$, eq.(\ref{O}). 
Away from the conformal point 
$\epsilon_0$, the perturbed WZNW model is no longer conformal.
 However, it is easy to see 
that in the limit $k\to-\infty$, eq.(\ref{beta})
 has a non-trivial fixed point 
\begin{equation}
\epsilon^*=-{2c_V(G)\over\pi k},\label{*}\end{equation}
at which the perturbed non-unitary WZNW model becomes a unitary WZNW
model with level $|k|-
2c_V(G)$ \cite{Soloviev1}. 
This can be checked by computing anomalous conformal dimensions 
and the Virasoro central charge.

However, the flow from $\epsilon_0$ to $\epsilon^{*}$ must be very
peculiar.  Indeed, in the 
middle of the flow, the parameter $\epsilon$ has to pass another point
\begin{equation}
\epsilon_{WZ}=-{c_V(G)\over\pi k},\label{middle}\end{equation}
at which the deformation kills the sigma-model kinetic term and
 the resulting theory has only the Wess-Zumino term, i.e.
\begin{equation}
S(\epsilon_{WZ})\sim\Gamma.\label{WZ}\end{equation}

This point does not appear to be special from the point of view of 
the beta function. For example, it is not a critical point.
 However, one can expect some very dramatic events 
happening as $\epsilon(t)$ approaches $\epsilon_{WZ}$,
 since the theory becomes purely 
topological. The properties of this theory are quite
 puzzling \cite{Stern} which make fairly 
difficult its quantum interpretation. In spite of certain pathology, 
the described perturbed 
WZNW model seems to mimic well the behaviour of
 Polyakov's confining string. Indeed, we can 
identify the coordinates $x^\mu$ from ansatz (\ref{ansatz})
 with the coordinates on the group manifold $G$. 
Then the variable tension will be given as follows
\begin{equation}
a(t)\sim k^2\left({1\over4\pi k}~+~
{\epsilon(t)\over4c_V(G)}\right).
\label{run}
\end{equation}

In order to make the link between the perturbed WZNW model
 and Polyakov's ansatz more 
precise, we have to couple the former to 2D gravity.
 We shall show that this coupling will 
fix the pathologies of the original theory (\ref{perturbation}).

The point to be made is that the Virasoro central
 charge of the unperturbed WZNW model is given as follows
 \cite{Knizhnik}
\begin{equation}
c_{WZNW}(G,k)={k\dim G\over k + c_V(G)}
\label{virasoro}
\end{equation}
and it is not necessarily equal to 26. 
Therefore, the given WZNW model has to be thought of 
as being a certain non-critical string. 
It is well-known that a non-critical string in $D$ 
dimensions with coordinates $x^\mu$ can be viewed 
as a critical string in the ($D+1$)-
dimensional space formed by $X^M=(x^\mu,\;\phi)$, 
with $\phi$ being the Liouville field. 

As a result of the gravitational Ward identities, the interaction with
2D gravity  makes any 
two-dimensional quantum field theory conformal. 
In particular, the perturbed WZNW model in question coupled to 2D
gravity  has to be a CFT. In other words, the gravitational interaction 
modifies the renormalization group properties of the system. 
The effects of the gravitational dressing have been studied in light-cone 
\cite{Polyakov2},\cite{Knizhnik2} and conformal gauges in \cite{David}
\cite{Distler1}, \cite{Klebanov} and  \cite{Schmidhuber} 
 (see also \cite{tks}, \cite{ag}, \cite{Schmidhuber2},
 \cite{dorn} and references therein).  In a light-cone gauge the 
 renormalization of a one-loop beta function, i.e. constant $C$
 was found 
  for marginal operators   \cite{Klebanov}
\begin{equation}
\bar{C} = \frac{\kappa + 2}{\kappa + 1}C = -\frac{2}{Q\alpha} C
\end{equation}
 and in conformal
 gauge the result  was obtained also for quasimarginal
 operators with bare dimensions $\Delta_0$ close to $1$ 
 \begin{equation}
\bar{C} = -\frac{2}{\alpha(Q-2\alpha_1)} C
\end{equation}
 Let us note  that it  is still  an open
 problem to obtain the result for $\Delta_0 \neq 1$ in a light-cone gauge.
 For $c>25$ we have  
\begin{eqnarray}
 \kappa+2 =\frac{1}{12}\left (c-13 + \sqrt{(c-1)(c-25)}\right ),
Q = \sqrt{{|c-25|\over3}}, ~~~~~~~~
\nonumber \\
\alpha = -{Q\over2}~+~{1\over2}\sqrt{Q^2+8}, ~~~~~~~
\alpha_1=-{Q\over2}~+~\sqrt{{Q^2\over4}-(2\Delta_0-2)},
\label{central}
\end{eqnarray}
The gravitationally dressed beta function is given by
\begin{equation}
\bar\beta\equiv{d\epsilon\over d\ln \Lambda}=
(2-2\Delta)\epsilon~-~\pi\bar C\epsilon^2~+~...
\label{dressed}\end{equation}
 where  $\Delta$ is the KPZ  gravitational scaling dimension \cite{Knizhnik2}
\begin{equation}
 \Delta - \Delta_0 = \frac{\Delta(\Delta-1)}{\kappa + 2},~~~~
 1-\Delta = {\alpha_1\over\alpha}
\label{definitions}
\end{equation}
 
 One can see   that there is a  non-trivial IR fixed point
\begin{equation}
\epsilon^{*}=-{\alpha_1(Q-2\alpha_1)\over\pi}
\label{fixedpoint}
\end{equation}
  One  can trust this perturbative result only if $|\epsilon^{*}| <<1$
  which is true in a large $k$ limit as will be shown later.

In order to find the Liouville field $\phi$ dependence 
 of  the gravitationally dressed $\beta$-function,
we need a precise definition of the physical scale. 
 In \cite{Schmidhuber}, \cite{Klebanov} it was suggested
  to identify the scale $\Lambda^{-2}$
with the cosmological constant
operator, also known as the ``tachyon background'' $T(\phi)$.
 Only its leading asymptotic for large $\phi$ is known
$ T(\phi)\sim e^{-\alpha\phi}$ and
 in this paper we  are not going to address the issue of
 the intermediate $\phi$ behaviour. It is quite possible
 that $ T(\phi)$ becomes zero at  some finite $\phi$ and this
 will be the IR endpoint of the RG flow.

The well-known problem with the non-critical strings is that
 we do not know how to quantize them when the Virasoro central
 charge lies between 1 and 25. Therefore, in order to move 
forward with the theory in question, we have to make sure that
 its Virasoro central charge $\ge25$. 
Otherwise, it might be difficult to construct any more or less 
realistic model of 
the confining string. As we shall see, our choice will be fully justified.

We can also assume that the zigzag-invariant string is confined to a
low-dimensional 
domain of a larger space, that is there are extra 
coordinates $y^n$ which the zigzag-invariant
string does not see. In other words, we assume
 that the Virasoro central charge without gravity has the following structure
$c=c_{WZNW}~+~c_y$
where $c_y$ is the Virasoro central charge
 associated with the additional coordinates $y^n$.
The perturbation will be carried out only in the $x$-sector, 
leaving $y^n$ untouched.

In the limit $k\to-\infty$, the WZNW-Virasoro central
 charge has the following expansion in $1/k$
\begin{equation}
c_{WZNW}(G,k)=\dim G~+~{c_V(G)\dim G\over |k|}~+~{\cal 
O}(1/k^2).\label{expansion}\end{equation}
Thus, if we choose $\dim G$ and $c_y$ such that
$\dim G~+~c_y=25$
then we shall have a non-critical string whose Virasoro central charge
is just slightly greater than 25. 
This condition will allow us to study the gravitational dressing using the 
$1/k$-expansion method.

Within the described setup, we obtain the following asymptotic expressions:
\begin{eqnarray}
Q&=&\sqrt{{c_V(G)\dim G\over3|k|}}~+~...\nonumber\\ & & \\
\alpha_1&=&\sqrt{{c_V(G)\over|k|}}\left[{-\sqrt{{\dim G\over3}}~+~\sqrt{{\dim 
G\over3}~+~8}\over2}\right]~+~...,
\nonumber\label{results}\end{eqnarray}
where dots stand for higher in $1/k$-corrections.

After substitution of the above formulas into eq.(\ref{fixedpoint}), we
derive the 
following solution for the non-trivial fixed point in the presence of gravity
\begin{equation}
\epsilon^{*}= -{c_V(G)\over\pi|k|}\left({-\sqrt{{\dim
G\over3}}~+~\sqrt{{\dim G\over3}~+~8}\over2}\right)
\left(2{\sqrt{\dim G\over3}}~-~\sqrt{{\dim 
G\over3}~+~8}\right).\label{IR2}\end{equation}
One can see that the location of this critical point depends on the
group $G$.  In order for 
our theory to make a contact with Polyakov's confining string,
 we have to impose the 
following condition
\begin{equation}
\epsilon^{*}=\epsilon_{WZ}={c_V(G)\over\pi|k|},\label{want}\end{equation}
which guarantees that the perturbed CFT coupled to 2D gravity runs to
the zigzag-invariant background and settles down there. 
Eq.(\ref{want}) results  in an equation for $\dim G$
\begin{equation}
(\sqrt{x^2+8}-2x)(\sqrt{x^2+8}-x) = 2,~~~
x^2 = {\dim G\over3}
\end{equation}
  and nobody  promised us that the solution will  give us a positive
 integer  $\dim G$.
 Amusingly enough  the  solution  is
\begin{equation}
x^2 = 1, ~~\dim G =3.
\label{dimG=3}
\end{equation}
Since the group $G$ has to be noncompact 
(because $k$ is negative), we end up with a unique solution
\begin{equation}
G = SL(2).\label{sl2}\end{equation}
Correspondingly, we find that
$c_y=22$. Because, the perturbation did not affect the $y$-sector,
 the latter stays completely 
factorized from both $x^\nu$ and the Liouville field throughout the flow.

In the limit $k\to-\infty$, our non-critical string with 
$G=SL(2)$ and $c_y=22$ at the UV-
critical point will flow to the fixed point at which
 the $x$-sector of the theory acquires 
the zigzag symmetry. It seems to be appropriate to call
eq.(\ref{want}) the zigzag-anomaly 
free condition. The fact, that the zigzag symmetry fixes
 the group $G$ is not completely 
strange. We could have expected that the zigzag symmetry
 should have an anomaly at the 
quantum level (if we have started with a zigzag invariant classical
model). In string theory, 
the anomaly cancellation always results in certain 
restrictions on the string background.

Condition (\ref{sl2}) does not necessarily mean that the bosonic 
zigzag-invariant string can 
exist only in the three-dimensional space formed by the 
renormalization flow of the WZNW 
model on $SL(2)$ coupled to 2D gravity. The zigzag symmetry 
also can be realized via 
deformations of the $SL(2)/U(1)$ coset \cite{Soloviev3}
 which will lead to a certain two-
dimensional zigzag-invariant string (see appendix). 
The $U(1)$-gauge invariant perturbation 
operator $\tilde O$ is defined as follows \cite{Soloviev3}
\begin{equation}
\tilde O={1\over c_V(G)}\left[J^a\bar J^b\phi^{ab}+(1+{4\over k})J^a\bar 
J\phi^{a3}+(1+{4\over k})J\bar J^b\phi^{3b}+(1+{4\over k})^2 J
\bar J\phi^{33}\right],\label{U(1)}\end{equation}
where $J$ is the affine current associated with the subgroup $U(1)$.

The gateway to higher dimension formulations of the zigzag-invariant
 string opens up via 
tensoring of the WZNW model on the non-compact $SL(2)$ al level $k$
with the WZNW model on the compact $SU(2)$ at level $l$. 
It turns out that the level-dependence in the sum of the 
Virasoro central charges of these two WZNW models can vanish.
 Indeed, the sum is given as 
follows
\begin{equation}
c_{sum}={3k\over k+2}~+~{3l\over l+2}=
6~-~6\left({1\over\tilde k}~+~{1\over\tilde 
l}\right),\label{tensoring}\end{equation}
where we made the following change of the levels
$k=\tilde k~-~2,~
l = \tilde l~-~2$. So if we take $\tilde l=-\tilde k$ then 
$c_{sum}=6$ 
and does not depend on the levels. Obviously, $k$ has to be negative 
integer in order for the WZNW model on $SU(2)$ to be well defined.

Thus, via tensoring of one WZNW model on $SL(2)$ at level $k$ with the
two  WZNW models 
described above, we will get a system whose $Q$ has exactly
 the same $k$-dependence as for a 
single WZNW model on $SL(2)$. 
This is sufficient for the deformed two WZNW models on the non-
compact groups to flow to the zigzag-invariant fixed point. 
In the case of two WZNW models on 
$SL(2)$, we obtain a six-dimensional zigzag-invariant string.
 The compact sector (formed by 
$y^m$) will have one $SU(2)$ and an additional 16-dimensional compact
space,  so that the 
whole target space, including the Liouville mode, will be
26-dimensional.  We can also 
consider tensoring cosets. This way we can derive zigzag-invariant
 strings in dimensions from  $D=2$ to $D=23$, though the total number
 of the target space dimensions will be always 26. 
The dimension $D=23$ is maximal, since the compensating WZNW model
 (or its coset) on $SU(2)$  will have dimension, at least, equal to 2,
which will be the minimal dimension of the compact  space.

\section{The zigzag-invariant string with the world-sheet supersymmetry}
\setcounter{equation}{0}

The philosophy of the world-sheet formulation of the supersymmetric
 zigzag-invariant string 
almost completely repeats the bosonic case. 
The differences appear only in numerical details 
which, however, will be shown to lead us to quite 
remarkable consequences compared to the 
purely bosonic formulation. We restrict our consideration to 
the case of $N=1$ supersymmetry.

It is convenient to use the $N=1$ superspace formulation of the 
WZNW model \cite{Di Vecchia}. The $N=1$ supergravity dressing 
has  been studied in \cite{Polyakov3},\cite{Distler2},\cite{Grisaru}
 and here we shall skip details which could be found in this
publications.
The gravitationally dressed beta function is given by the
 expression  similar to  (\ref{dressed})
\begin{equation}
\hat\beta\equiv{d\epsilon\over d\ln \Lambda}=
 2\frac{{\hat\alpha_1}}{{\hat\alpha}}\epsilon~+~
{2\pi\over \hat Q-2\hat\alpha_1}\epsilon^2~+~...
\label{SUSY}
\end{equation}
 and the  IR fixed point is 
$\epsilon^{*}=-{\hat\alpha_1}({\hat Q}-2{\hat\alpha_1})/\pi$.
 Here  $\hat\alpha_1$ and $\hat Q$ are 
supersymmetric counterparts of the bosonic $\alpha_1$ and $Q$.
 They are given as follows
\begin{eqnarray}
\label{new}
\hat Q=\sqrt{{\hat c-9\over2}}~~~~~~~
\hat\alpha_1=-{\hat Q\over2}~-~\sqrt{{\hat Q^2\over4}-2(\hat\Delta_0-
{1\over2})}
\end{eqnarray}
where   $\hat c$ is the $N=1$ super-Virasoro central charge 
and bare anomalous dimension of a deformation operator is
\begin{equation}
\hat\Delta_0={1\over2}~+~{c_V(G))\over k}
\label{sup}
\end{equation}
 It is related to the Virasoro central charge $c$ of the component
theory as
\begin{equation}
\hat c ={2\over3}c
\label{supVir}
\end{equation}
The coefficient $2/3$ will turn out to be absolutely crucial.

We would like to consider a supersymmetrical CFT
 whose super-Virasoro central charge $\hat c$ 
is slightly greater than 9. This may require extra dimensions
 in the target  space in addition to the supersymmetrical WZNW model
on $G$. Thus, as in the bosonic case, the target space 
will consist of the $x$-sector and the $y$-sector, where $x^\mu$ and
$y^n$  are now $N=1$ superfields. 
The dimensions of the given sectors are related as follows
$\dim G~+~\hat c_y=9$.
In the limit $k\to-\infty$, we find
\begin{equation}
\hat c - 9={2\over3}c - 9= {2c_V(G)\dim G\over3|k|}
\label{dif}
\end{equation}
which is, in fact, an exact expression (without higher in $1/k$-terms).
Substitution of the last expression and $\hat\Delta_0$ into (\ref{new}) 
gives rise to 
exactly  the same expressions for $\hat\alpha_1$ and $\hat Q$ as in bosonic 
eqs.(\ref{results}) due to the factor $2/3$ in formula (\ref{supVir}),
 which means the same value for $\epsilon^{*}$. 
Therefore, in the 
$N=1$ supersymmetric case, we shall arrive at the same zigzag anomaly 
cancellation condition  as in (\ref{dimG=3}), 
that is the group $G$ still has to be $SL(2)$. This allows us to 
construct a three-dimensional zigzag-invariant string with $N=1$ 
supersymmetry on the world-sheet. 
It is also straightforward to obtain a two-dimensional $N=1$ zigzag-
invariant string based on the super-coset $SL(2)/U(1)$.

The advantage of the supersymmetric case compared to the bosonic one 
is that the $N=1$ supersymmetry lowers the number of possible $N=1$ 
supersymmetrical zigzag-invariant strings 
in higher dimensions. Indeed, let us consider two $N=1$ 
supersymmetric WZNW models defined on 
$SL(2)$ at level $k$ and $SU(2)$ at level $l$, correspondingly. 
Remarkably, the sum of their super-Virasoro central charges does not 
depend on the Kac-Moody level, if $l=-k$
\begin{equation}
{\hat c}(SL(2),k)~+~{\hat
c}(SU(2),-k)={2\over3}\left[{3(k-2)\over(k-2)+2}
~+~{3(-k-2)\over(-k-2)+2}~+~3\right]=6
\label{6}
\end{equation}
Because of this cancellation, we can tensor these two WZNW models 
(or their gauged versions) 
with one $N=1$ supersymmetric WZNW model on $SL(2)$ (or the corresponding 
coset) at the large negative level $k$. 
Since $k$ is related to the Kac-Moody level of the WZNW model on a 
compact group, it has to be negative integer. However, this condition does 
not affect our computations. This level-cancellation mechanism is
similar to the procedure considered in the 
bosonic case. However, in the supersymmetric case, the Kac-Moody
levels come up already shifted by the right number.

Let us describe how one can obtain a $D=4$ zigzag-invariant string. 
In order to get a four-dimensional space formed by $x^\mu$, 
we have to take two super-cosets $SL(2)\over U(1)$ at 
the same level $k$. Each of the given cosets is deformed by the 
corresponding gauge invariant operator $\hat{\tilde O}$. 
The gauge invariant supersymmetric operator $\hat{\tilde O}$ is 
constructed as the $N=1$ superspace generalization of the bosonic operator 
$\tilde O$ in eq.(\ref{U(1)}). 
The anomalous conformal dimension of the super-operator $\hat{\tilde O}$ 
coincides with $\hat\Delta_0$ in eq.(\ref{sup}). In order to have the
 same $k$-dependence in the super-Virasoro central charge, we can
tensor the given two super-cosets with one super-WZNW model on $SU(2)$
at level $l=-k$. This means that $k$ must be negative integer. This 
compact space will be a part of the five-dimensional compact 
space $\Sigma^5$ formed by $y^n$.
 Since $SU(2)$ is a three-dimensional space, the $\Sigma^5$-compact 
space has to have the following structure
\begin{equation}
\Sigma^5=SU(2)\times\Sigma^2,\label{Sigma5}\end{equation}
where $\Sigma^2$ is a two-dimensional compact space with
$\hat c_{\Sigma^2}=2$.
The compact space $\Sigma^2$ can be realized either as a torus or a sphere. 
So the target space of the given string model at the UV-point will 
be $M^D\times SU(2)\times 
\Sigma^2\times\mbox{Super-Liouville}$.

There is another interesting solution for $\Sigma^5$. Instead of 
$SU(2)\times\Sigma^2$, we can take the following coset
\begin{equation}
T^{1,1}={SU(2)\times SU(2)\over U(1)}
\label{Einstein}
\end{equation}
If we take the Kac-Moody level of each of these two $SU(2)$s being equal to 
$-2k$, then we get the right $k$-dependence in the Virasoro central
charge. It is worth mentioning that this space has the symmetry of 
the 5-dimensional Einstein space \cite{Romans} recently discussed in the 
context of the AdS/CFT-duality \cite{Gubser2}. 
This fact seems to be quite intriguing.

The similar procedure allows us to construct zigzag-invariant strings in 
$D=5$, $D=6$ and $D=7$ dimensions. 
Also it is quite curious that the mentioned above tensoring does not permit 
us to have $M^D$ with $D>7$. The dimension seven emerges as the maximally 
possible dimension of the $N=1$ zigzag-invariant string
\footnote{It might be interesting to see whether this 
restriction on the dimensionality of $M^D$ has anything to do with the 
restriction on the dimensionality of simple super-anti-De Sitter
supergroups,  $d\le7$ \cite{Gunaydin}.}. 
Indeed, in order to get $D=8$, we would have to take two $SL(2)$ and one 
$SL(2)\over U(1)$. To get the right dependence on $k$ in the 
super-Virasoro central charge, we then would need 
at least one $SU(2)$ in the $y^n$-sector. However, the latter has to have 
$\hat c_y\le1$, which would be impossible to get.

All in all we can draw  Table~1 for $N=1$ supersymmetric CFTs which lead us 
to the $N=1$ 
zigzag-invariant strings in $D$ dimensions.

\begin{table}\centering
\begin{tabular}{|l|l|l|}
\cline{1-3}
\vbox to1.79ex{\vspace{1pt}\vfil\hbox to10.67ex{\hfil $D$\hfil}} & 
\vbox to1.79ex{\vspace{1pt}\vfil\hbox to43.17ex{\hfil $M^D$\hfil}} & 
\vbox to1.79ex{\vspace{1pt}\vfil\hbox to35.00ex{\hfil $\Sigma^{9-D}$\hfil}} 
\\

\cline{1-3}
\vbox to1.50ex{\vspace{1pt}\vfil\hbox to10.67ex{\hfil 2\hfil}} & 
\vbox to1.50ex{\vspace{1pt}\vfil\hbox to43.17ex{\hfil $SL(2)/U(1)$\hfil}} & 
\vbox to1.50ex{\vspace{1pt}\vfil\hbox to35.00ex{\hfil $\Sigma^7$\hfil}} \\

\cline{1-3}
\vbox to1.79ex{\vspace{1pt}\vfil\hbox to10.67ex{\hfil 3\hfil}} & 
\vbox to1.79ex{\vspace{1pt}\vfil\hbox to43.17ex{\hfil $SL(2)$\hfil}} & 
\vbox to1.79ex{\vspace{1pt}\vfil\hbox to35.00ex{\hfil $\Sigma^6$\hfil}} \\

\cline{1-3}
\vbox to4.50ex{\vspace{1pt}\vfil\hbox to10.67ex{\hfil 4\hfil}} & 
\vbox to4.50ex{\vspace{1pt}\vfil\hbox to43.17ex{\hfil ${SL(2)\over 
U(1)}\times{SL(2)\over 
U(1)}$\hfil}} & 
\vbox to4.50ex{\vspace{1pt}\vfil\hbox to35.00ex{\hfil 
$SU(2)\times\Sigma^2,~T^{1,1}$\hfil}} 
\\

\cline{1-3}
\vbox to4.50ex{\vspace{1pt}\vfil\hbox to10.67ex{\hfil 5\hfil}} & 
\vbox to4.50ex{\vspace{1pt}\vfil\hbox to43.17ex{\hfil 
$SL(2)\times{SL(2)\over U(1)}$\hfil}} & 
\vbox to4.50ex{\vspace{1pt}\vfil\hbox to35.00ex{\hfil 
$SU(2)\times\Sigma^1,~\left[{SU(2)\over U(1)}\right]^2$\hfil}} \\

\cline{1-3}
\vbox to4.50ex{\vspace{1pt}\vfil\hbox to10.67ex{\hfil 6\hfil}} & 
\vbox to4.50ex{\vspace{1pt}\vfil\hbox to43.17ex{\hfil $SL(2)\times 
SL(2),~\left[{SL(2)\over U(1)}\right]^3$\hfil}} & 
\vbox to4.50ex{\vspace{1pt}\vfil\hbox to35.00ex{\hfil $SU(2)$\hfil}} \\

\cline{1-3}
\vbox to4.50ex{\vspace{1pt}\vfil\hbox to10.67ex{\hfil 7\hfil}} & 
\vbox to4.50ex{\vspace{1pt}\vfil\hbox to43.17ex{\hfil 
$SL(2)\times\left[{SL(2)\over U(1)}\right]^2$\hfil}} & 
\vbox to4.50ex{\vspace{1pt}\vfil\hbox to35.00ex{\hfil ${SU(2)\over 
U(1)}$\hfil}} \\

\cline{1-3}
\end{tabular}

\caption{$M^D\times\Sigma^{9-D}\times$(Super)Liouville-geometry of the
target space before deformation. Here $M^D$ is the space which is 
deformed by the operator $\hat O$, one for each 
factor in case when $M^D$ is a product, 
$\Sigma^{9-D}$ is the compact space formed by $y^m$.}
\end{table}

It is interesting to point out that all $M^D\times\Sigma^{9-D}
\times\mbox{Super-Liouville}$ 
spaces are ten-dimensional. In other words, the deformed super-CFT coupled to 
$N=1$ supergravity describes certain ten-dimensional non-trivial
 backgrounds of the critical superstring. 
All these backgrounds have to be solutions of ten-dimensional supergravity 
compactified on various spaces. Therefore, we may expect that this
theory is unitary, despite 
the fact that we consider non-unitary WZNW models. The signature of $M^2$, 
$M^4$, $M^5$, $M^6$ and $M^7$ depends on whether we gauge away the
compact or non-compact $U(1)$ subgroup of $SL(2)$.

\section{Conclusion}

By using the method of perturbed CFT, we have managed 
to give the world-sheet formulation of 
the zigzag-invariant string in various dimensions. 
Within our approach, the zigzag symmetry 
imposes very strict conditions on the geometry and dimensionality
 of the target space. We considered here only one charge deformation
 - it will be interesting to study the case when model is deformed
 by several relevant operators. We did not discuss here why the
$SL(2)$ group manifold plays such a special role and is there any
connection between this SL(2) and hidden SL(2) in the Liouville
theory, is there an enhancement of the target space symmetry at the
 horizon, what are  the
  correlation functions of  deformed model, as
well as many other important   questions which we are planning
 to discuss in further publications. 
 We do not claim also  that our  approach based on  a gravitational
  dressing of perturbed world-sheet CFT covers all possible
 string solutions consistent  with the zigzag symmetry. 
It might be the case that there are some more non-trivial zigzag-
invariant string backgrounds (maybe with $D>7$) 
which can not be studied within perturbation 
theory. 
 Nevertheless we think that  this  approach
 can be useful for understanding the 
relation between gauge field theories and string theory.

\section{Acknowledgments}
 I.I.K. thanks A.M. Polyakov for stimulating discussion about
 confining strings.  O.S. would like to thank PPARC for financial support.

\section{Appendix}
\setcounter{equation}{0}

In this appendix we would like to discuss a possibility of constructing 
zigzag-invariant 
strings based on deformations of a generic coset construction $G/H$. The 
procedure consists of two steps. The first one is to take into account
 the gauge dressing of the kinetic term of 
the WZNW model on $G$ with respect to the gauge group $H$ \cite{Lewis}.
 The second step is to take into account the gravitational dressing of
the deformed gauged WZNW model.

Within the $1/k$-method, we find the gauge-dressed anomalous conformal 
dimension of the kinetic term
\begin{equation}
\Delta_{g.d.}=1~-~{c_V(G)-c_V(H)\over|k|}~+~...
\label{g.d.}
\end{equation}
Correspondingly, the Virasoro central charge of the $G/H$ coset is given as 
follows
\begin{equation}
c_{G/H}=\dim G~-~\dim H~+~{c_V(G)\dim G\over|k|}~-~{c_V(H)\dim 
H\over|k|}~+~...\label{G/H}
\end{equation}

Taking into account the gravitational dressing gives rise to the following
zigzag-anomaly cancellation condition
\begin{equation}
c_V(G)(\dim G - 3)=c_V(H)(\dim H -3).
\label{condition}
\end{equation}
In case of  ordinary Lie algebras, this equation has only one solution. 
Namely,
\begin{eqnarray}
\dim G=3 ~~~~~~~~~c_V(H)=0,
\label{onesol}
\end{eqnarray}
which is nothing but $SL(2)/U(1)$ which we have already discussed in the 
paper. However, in case of super-algebras \cite{Kac}, there exist 
many other solutions of the type $G/SL(2)$, when $G$ is a supergroup
 with $c_V(G)=0$ and $SL(2)$ is an ordinary Lie group. It migh be 
interesting to look for solutions similar to the supergeometry 
considered in \cite{Metsaev}.

\end{document}